\documentclass{article}

\usepackage{arxiv}
\usepackage[utf8]{inputenc}
\usepackage[T1]{fontenc}
\usepackage{hyperref}
\usepackage{url}
\usepackage{booktabs}
\usepackage{amsfonts}
\usepackage{amsmath}
\usepackage{amssymb}
\usepackage{nicefrac}
\usepackage{microtype}
\usepackage{graphicx}
\usepackage{array}
\usepackage{tabularx}
\usepackage{longtable}
\usepackage{xcolor}
\usepackage{setspace}
\usepackage{enumitem}

\graphicspath{{./images/}}

\title{Temporal Coverage Bias in Financial Panel Data: A Coverage-Aware
Structuring Framework with Evidence from the Dhaka Stock Exchange}

\author{
  Tashreef Muhammad \\
  Department of Computer Science and Engineering \\
  Southeast University \\
  Dhaka, Bangladesh \\
  \texttt{tashreef.muhammad@seu.edu.bd}
}

\begin{document}
\maketitle

\begin{abstract}
A common practice in empirical finance is to construct calendar-aligned panels that implicitly treat all instruments as having existed for the full observation period. When securities with different listing histories are combined into such panels without explicit coverage constraints, price histories can be inadvertently extended before valid trading ever began. This paper formalizes this problem and proposes a coverage-aware structuring framework built around instrument-level observation windows encoded through structured metadata and an availability matrix. Applied to end-of-day data from the Dhaka Stock Exchange spanning October 2012 to January 2026 and covering 486 instruments across equities, mutual funds, and fixed-income securities, the framework reveals substantial distortions from naive temporal alignment. Controlled ARIMA-based experiments establish the mechanism through which padded observations corrupt return dynamics, and volatility analysis across 53 instruments - combining return variance decomposition with GARCH-based unconditional variance estimation - shows that naive calendar alignment suppresses return volatility by roughly 20\% on average, with GARCH unconditional variance distortions exceeding 26\% and the effect present in more than 90\% of the instruments examined. These figures are derived from the milder of the two naive constructions studied, representing a lower bound on the full severity of the bias. The distortion is not confined to volatility estimation: any analytical method that requires calendar alignment of heterogeneous instrument histories is susceptible, including pairwise similarity measures such as dynamic time warping, covariance-based portfolio construction, factor model regression, co-integration testing, and panel-level feature representations used in the fine-tuning of temporal foundation models. The coverage-aware framework and availability matrix introduced here provide a reusable correction applicable across all of these settings. Although demonstrated on financial data, the framework applies to any panel combining entities with heterogeneous entry dates, including sensor networks, longitudinal clinical datasets, country-level economic panels, and environmental monitoring arrays. Instrument listing coverage, in short, is not a minor preprocessing consideration but a first-order variable in panel construction, with direct consequences for volatility forecasting, risk management, and asset pricing research.

\end{abstract}


\section{Introduction}

Financial time-series datasets support a wide range of empirical work in asset pricing, volatility modeling, market microstructure, and risk management \cite{campbell1997econometrics, tsay2005analysis, bekaert2003emerging, fama1991efficient}. The typical approach is to combine individual instrument histories into a unified, calendar-aligned panel, which makes cross-sectional comparisons straightforward and allows standard econometric models to operate on rectangular data structures \cite{campbell1997econometrics, fama1993common, cochrane2005asset}. This is a sensible and widely adopted convention, but it carries an assumption that is rarely stated explicitly: that every instrument in the panel existed for the full observation window.

In practice, this assumption is almost never satisfied. Securities are listed at different times, and newer instruments naturally have shorter histories. When a dataset is aligned to a global calendar without recording each instrument's listing interval, the alignment process can implicitly extend a security's price history before its first trading date. Standard preprocessing steps - forward filling, backward filling, or simple panel construction routines - can all produce this effect without any explicit decision being made \cite{lopezdeprado2018advances}.

The consequences are not trivial. Financial models are sensitive to the statistical properties of return series, and spurious pre-listing returns alter those properties in ways that are difficult to detect after the fact \cite{cont2001empirical}. Volatility estimation in particular is vulnerable, given how heavily portfolio construction, derivative pricing, and risk modeling rely on accurate variance estimates \cite{poon2003forecasting, hansen2005forecast}. A distortion embedded in the data construction stage propagates quietly through every downstream analysis that uses those data.

This problem is related to survivorship bias - the well-documented tendency for datasets that include only currently active instruments to overstate historical returns \cite{brown1992survivorship, malkiel1995returns} - but it is mechanistically distinct. Survivorship bias works by \emph{exclusion}: instruments that failed are dropped, inflating apparent performance \cite{brown1992survivorship, brown1995survival, carhart1997persistence}. The problem studied here works by \emph{extension}: instruments that were not yet listed are treated as though they were, which dilutes return variability rather than inflating it. Other recognized data integrity problems include data-snooping bias \cite{lo1990datasnooping} and look-ahead bias \cite{lopezdeprado2018advances}, but this particular distortion - arising specifically from backward extension before an instrument's valid listing date - has not, to the best of the author's knowledge, been formally named or characterized in the literature.

This paper addresses that gap using end-of-day trading data from the Dhaka Stock Exchange (DSE), an emerging market that has been studied through both GARCH-based volatility models and deep learning approaches \cite{hasan2017volatility, siddikee2016volatility, muhammad2023transformer}. A coverage-aware structuring framework is developed that encodes instrument availability explicitly through structured metadata and an availability matrix, making the valid observation window of each instrument a first-class component of the dataset rather than an implicit assumption. Although the empirical work is grounded in DSE data, the underlying problem applies to any financial panel constructed without explicit coverage encoding.

The dataset covers October 2012 to January 2026. Using a sample of 53 instruments with heterogeneous listing histories, the analysis first uses ARIMA models to illustrate the mechanism through which padding corrupts return dynamics, then conducts a large-scale robustness analysis using GARCH-based unconditional variance estimation - the primary volatility measure - to quantify the distortion across coverage-aware and naive dataset constructions.

Three contributions follow from this work.

\begin{enumerate}[label = \roman*)]
  \item \textbf{Coverage-aware structuring of financial time series.} A formal dataset representation is introduced in which each instrument's observation window is encoded explicitly through metadata and a binary availability matrix. This prevents price histories from being silently extended before valid listing dates and provides a reusable foundation for building clean financial panels.

  \item \textbf{Formalization of temporal coverage bias.} The term \emph{temporal coverage bias} is defined to describe the distortion that arises when heterogeneous listing intervals are ignored during calendar alignment. The bias is shown to be structured and predictable, not random noise, and is distinguished clearly from related but distinct problems such as survivorship bias and look-ahead contamination.

  \item \textbf{Empirical quantification of volatility distortion.} Using DSE data from October 2012 to January 2026, the analysis quantifies distortion under two naive constructions of increasing severity. Under calendar-day forward-filling - the milder and more common construction - return volatility is suppressed by roughly 20\% on average, GARCH unconditional variance distortions exceed 26\%, and the effect is present in more than 90\% of the 53 instruments in the cross-instrument sample; these figures represent a lower bound on the full severity. Under backward extension to the global panel start - the more severe construction - return volatility suppression reaches 36.6\% for a single-instrument illustration, and applying it across all 53 instruments causes GARCH estimation to break down entirely in 22 cases ($\hat{\alpha} + \hat{\beta} \geq 0.999$), underscoring that pre-listing backward extension does not merely distort volatility estimates but can render conditional variance modelling infeasible. The underlying dataset, covering 486 DSE instruments with full coverage metadata and availability matrices, is publicly available on Mendeley Data \cite{muhammad2026dhaka}.
\end{enumerate}

\paragraph{Scope of affected analyses.}
While this paper quantifies the effect through volatility estimation, the root cause --- calendar alignment of heterogeneous instrument histories --- is a data construction problem that sits upstream of any modelling choice. Temporal coverage bias therefore affects any analytical method that requires a common time axis across instruments. This includes, but is not limited to:

\begin{itemize}
  \item \textbf{Pairwise similarity measures} (e.g., dynamic time warping, Pearson correlation, covariance matrices): pre-listing zero-return sequences create phantom similarity between recently-listed instruments and inflate or suppress pairwise distances, corrupting the relational structure that clustering and portfolio optimisation depend on.

  \item \textbf{Factor model regression}: Fama--French style cross-sectional regressions and time-series factor loadings are estimated on panel returns; including instruments with fabricated pre-listing observations biases factor exposures and residual variance estimates.

  \item \textbf{Co-integration and vector autoregression}: tests for long-run relationships between instruments require accurate joint return histories; backward-extended series introduce spurious stationarity in the padding segment that can mask or manufacture cointegrating relationships.

  \item \textbf{Portfolio construction}: mean--variance optimisation, risk parity, and minimum-variance portfolios rely on covariance estimates; suppressed volatility for recently-listed instruments causes them to be systematically over-weighted.

  \item \textbf{Backtesting and event studies}: any backtest or event study that draws on a calendar-aligned panel without coverage encoding will silently include pre-listing periods as valid observation windows, inflating apparent sample sizes and distorting pre-event baselines.

  \item \textbf{Machine learning and foundation model fine-tuning}: models trained or fine-tuned on panel-formatted data (including temporal foundation models that process multiple instruments on a common time axis) will learn from fabricated observations as though they were genuine signal, degrading learned representations in proportion to the fraction of recently-listed instruments in the training panel.
\end{itemize}

The correction is identical in all cases: restrict analysis to each instrument's valid observation window as encoded in the availability matrix. The coverage-aware framework introduced here is designed to make this restriction a default property of the dataset rather than a case-by-case analytical decision.

Although this paper demonstrates the problem and the correction in a financial markets setting, the underlying issue is not specific to finance. Temporal coverage bias is a panel data construction problem: it arises whenever entities with heterogeneous observation start dates are combined into a calendar-aligned panel without explicit coverage encoding. The same structural distortion affects any domain where this condition holds. Sensor networks where nodes are added incrementally over time will exhibit phantom similarity between co-deployed sensors and genuine signal distortion for early-deployed ones in any distance-based or correlation-based analysis. Longitudinal clinical datasets where patients enroll at different points in the study window will produce biased baseline estimates and distorted treatment effect measurements if pre-enrollment periods are imputed rather than excluded. Country-level economic panels used in development research - where certain indicators were only tracked from a specific year - will suppress variance for recently-tracked countries when naive calendar alignment extends their histories backward. Environmental monitoring arrays, social media platform panels, and industrial IoT telemetry datasets are all structurally equivalent. The availability matrix representation and coverage-aware restriction introduced here transfer to these settings without modification: the entity type changes, but the correction - encode each entity's valid observation window and restrict analysis to it - is identical.

The paper proceeds as follows: Section~\ref{sec:2} reviews related work on volatility modeling, financial dataset construction, and temporal alignment. Section~\ref{sec:3} describes the data and the coverage-aware framework. Section~\ref{sec:4} sets out the empirical methodology. Section~\ref{sec:5} reports results. Section~\ref{sec:6} concludes.

\section{Related Literature}
\label{sec:2}

\subsection{Volatility Modeling in Financial Time Series}

Volatility sits at the center of much of financial econometrics, given its role in portfolio construction, options pricing, and risk assessment \cite{poon2003forecasting, hansen2005forecast}. Classical time-series approaches model return dynamics through autoregressive and moving-average structures; the ARIMA framework introduced by Box and Jenkins \cite{box2015time} remains a standard tool and has been applied to financial forecasting across a range of markets \cite{siami2019comparative, kobiela2022arima, banerjee2014forecasting, sunki2024time}.

A key empirical regularity in return data is volatility clustering --- large moves tend to be followed by large moves, and quiet periods cluster similarly \cite{cont2001empirical}. Engle's ARCH model \cite{engle1982autoregressive} was developed precisely to capture this feature, and Bollerslev's GARCH generalization \cite{bollerslev1986generalized} has since become the workhorse for conditional variance estimation in both developed and emerging market contexts \cite{poon2003forecasting, alfeus2025improving, ali2022modelling}. DSE-focused studies have confirmed that Bangladeshi equity returns display time-varying volatility and asymmetric GARCH effects consistent with the broader emerging market literature \cite{hasan2017volatility, siddikee2016volatility}.

More recent work has pushed in several directions: multivariate GARCH extensions, realized volatility measures, and hybrid models that combine econometric structure with machine learning \cite{tse2002multivariate, chun2025volatility, leber2025sentiment, pastpipatkul2025volatility}. Despite this progress, all of these methods share a common vulnerability --- their outputs are only as reliable as the return data they receive. If that data carries structural distortions from how the panel was built, the distortions carry through. That dependency is the motivating concern of the present study.

\subsection{Financial Time-Series Dataset Construction and
Preprocessing}

Most large-scale empirical work in finance depends on multi-instrument panels aligned to a common calendar \cite{campbell1997econometrics, fama1993common}. This structure is standard in asset pricing research, enabling the cross-sectional regressions and factor models that dominate the empirical literature \cite{cochrane2005asset, fama2015five, harvey2016and}.

Building such panels is not entirely straightforward. L\'{o}pez de Prado \cite{lopezdeprado2018advances} argues at length that data construction choices, alignment conventions, labeling schemes, and treatment of gaps are among the most consequential decisions in quantitative finance, and that many apparent model failures trace back to flawed data rather than flawed models. The concern most relevant here is what happens when instruments with different listing intervals are forced into a common calendar grid. Unless listing dates are explicitly tracked, the alignment process implicitly treats each instrument as having existed since the panel start date. From that point, any downstream model that uses the panel will operate on data it takes to be valid but which, for recently listed instruments, contains no genuine market information at all \cite{lopezdeprado2018advances}.

The downstream impact is broad. Calendar-aligned panels with untracked listing intervals contaminate every analysis that treats the panel as a source of ground truth: pairwise distance matrices used in DTW-based clustering will encode phantom similarity between recently-listed instruments; covariance matrices used in portfolio optimisation will understate the variance of newer instruments; factor model regressions will estimate loadings on a mix of genuine and fabricated returns; and any machine learning model fine-tuned on the panel will learn from synthetic observations as though they were valid signal. The coverage-aware framework introduced in Section~\ref{sec:3} addresses this at the data construction layer, providing a single correction that benefits all of these downstream analyses simultaneously.

\subsection{Temporal Alignment, Missing Observations, and Dataset Integrity}

The standard statistical treatment of missing data --- through interpolation, imputation, or exclusion \cite{little2019statistical, schafer1997analysis, chourib2025missing} --- assumes that the absence of an observation is a contingent fact about measurement, not a fundamental fact about the world. A sensor failed, a record was lost, a transmission dropped. In those settings, imputation is a reasonable response.

Pre-listing periods in financial data are categorically different. Before an instrument is listed, there is no price to observe. There is no measurement failure to correct for. Extending a price series backward into that period - even with a method as mild as carrying the first observation forward - manufactures information that has no empirical basis \cite{little2019statistical}. The return that such a fabricated series implies on the day of first listing is, technically, zero indefinitely, since the filled price never changes. This drags unconditional variance downward and distorts any model that uses the full series.

Survivorship bias is the more familiar data integrity issue in finance: datasets limited to currently active instruments systematically exclude the losers, biasing return estimates upward \cite{brown1992survivorship, malkiel1995returns, brown1995survival}. Temporal coverage bias runs in the opposite direction and through a different mechanism. Rather than excluding instruments, it \emph{extends} them, and the effect is suppression of estimated volatility rather than inflation of returns. Data-snooping bias \cite{lo1990datasnooping} and look-ahead bias \cite{lopezdeprado2018advances} are also distinct: the former concerns spurious discoveries from repeated testing, and the latter concerns the use of future information in historical model inputs. None of these, as far as can be determined from the existing literature, corresponds to or overlaps with the temporal coverage distortion characterized here.

\subsection{Gap Analysis}

Table~\ref{tab:literature_summary} maps representative studies against the specific gap this paper addresses. The pattern is consistent: existing volatility modeling work assumes clean, correctly structured data; existing work on data construction does not examine the statistical consequences of listing interval misalignment; and prior DSE studies treat their datasets as pre-validated inputs. The intersection of these three blind spots is where the present contribution sits.

\begin{table}[htbp]
  \centering
  \caption{Overview of representative studies related to financial time-series modeling, dataset construction, and DSE research. The \emph{Limitation} column identifies the gap each study leaves with respect to temporal coverage bias.}
  \label{tab:literature_summary}
  \setlength{\tabcolsep}{4pt}
  \begin{tabular}{p{0.14\linewidth} p{0.18\linewidth} p{0.17\linewidth}
                  p{0.18\linewidth} p{0.22\linewidth}}
    \toprule
    \textbf{Study} & \textbf{Dataset / Market} & \textbf{Methodology}
    & \textbf{Research Focus} & \textbf{Limitation of this study} \\
    \midrule
    Engle (1982) \cite{engle1982autoregressive}
    & UK inflation data & ARCH model
    & Conditional variance modeling
    & Dataset construction not examined \\ \\[1pt]

    Bollerslev (1986) \cite{bollerslev1986generalized}
    & Financial return series & GARCH model
    & Volatility clustering
    & Assumes correctly structured data \\ \\[1pt]

    Brown et al.\ (1992) \cite{brown1992survivorship}
    & US mutual funds & Performance analysis
    & Survivorship bias in datasets
    & Temporal listing coverage not addressed \\ \\[1pt]

    Poon \& Granger (2003) \cite{poon2003forecasting}
    & Multiple markets & Volatility model survey
    & Volatility forecasting review
    & Dataset alignment not examined \\ \\[1pt]

    Lo \& MacKinlay (1990) \cite{lo1990datasnooping}
    & US equity data & Statistical testing
    & Data-snooping bias
    & Distinct bias type; listing coverage not studied \\ \\[1pt]

    L\'{o}pez de Prado (2018) \cite{lopezdeprado2018advances}
    & Financial datasets & Financial ML
    & Data leakage and labeling
    & Listing interval coverage bias not addressed \\ \\[1pt]

    Hasan (2017) \cite{hasan2017volatility}
    & Dhaka Stock Exchange & GARCH
    & DSE volatility estimation
    & Dataset construction assumed pre-validated \\ \\[1pt]

    Siddikee \& Rahman (2016) \cite{siddikee2016volatility}
    & Dhaka Stock Exchange & GARCH
    & DSE volatility clustering
    & Dataset construction not examined \\ \\[1pt]

    Muhammad et al.\ (2023) \cite{muhammad2023transformer}
    & Dhaka Stock Exchange & Transformer deep learning
    & Stock price prediction
    & Dataset construction assumed pre-validated \\ \\[1pt]

    Alfeus et al.\ (2025) \cite{alfeus2025improving}
    & Emerging markets & Realised volatility
    & Volatility forecasting accuracy
    & Temporal alignment assumptions not examined \\
    \bottomrule
  \end{tabular}
\end{table}

Three specific gaps emerge from this review.

\begin{enumerate}[label = \roman*)]
  \item While survivorship bias \cite{brown1992survivorship, brown1995survival} and missing data \cite{little2019statistical} have received sustained attention, the distortion arising from naive calendar alignment of heterogeneously listed instruments has not been formally characterized or quantified.

  \item Methodological work on financial data construction \cite{lopezdeprado2018advances} does not address what happens statistically when instrument histories are extended beyond valid listing dates, despite the centrality of data quality to volatility inference \cite{cont2001empirical, poon2003forecasting}.

  \item Prior DSE volatility studies \cite{hasan2017volatility, siddikee2016volatility} model return dynamics without examining whether the underlying datasets carry construction-induced distortions.
\end{enumerate}

This paper addresses all three by introducing a coverage-aware structuring framework, giving the distortion a name and a formal definition, and quantifying its empirical magnitude across a diverse sample of DSE instruments.

\section{Data and Coverage-Aware Structuring}
\label{sec:3}

The dataset and all associated metadata are publicly available on Mendeley Data \cite{muhammad2026dhaka}.

\subsection{Data Source and Dataset Construction}

The empirical analysis draws on end-of-day (EoD) price records for instruments listed on the Dhaka Stock Exchange. Data were sourced from the Amarstock financial data platform, which aggregates historical trading records for DSE-listed securities from publicly available market data.

The dataset covers October 2012 to January 2026 and includes 486 instruments spanning equities, mutual funds, government securities, and other exchange-traded instruments. For each instrument, the record contains:

\begin{itemize}
  \item trading date
  \item opening price
  \item highest price
  \item lowest price
  \item closing price
  \item traded volume
\end{itemize}

Two versions of the dataset were maintained in parallel. The \emph{unadjusted} version preserves raw reported prices as they appear in the historical record. The \emph{adjusted} version incorporates corrections for corporate actions - stock splits, dividends, and similar events - where available. Both versions are organized as instrument-level files to allow independent analysis of individual securities and flexible panel construction.

Instrument-level metadata was added to support coverage analysis. These metadata record identifiers, inferred instrument types, and listing coverage information derived from each instrument's earliest and latest observed trading dates.

\begin{figure}[htbp]
  \centering
  \includegraphics[width=0.75\linewidth]{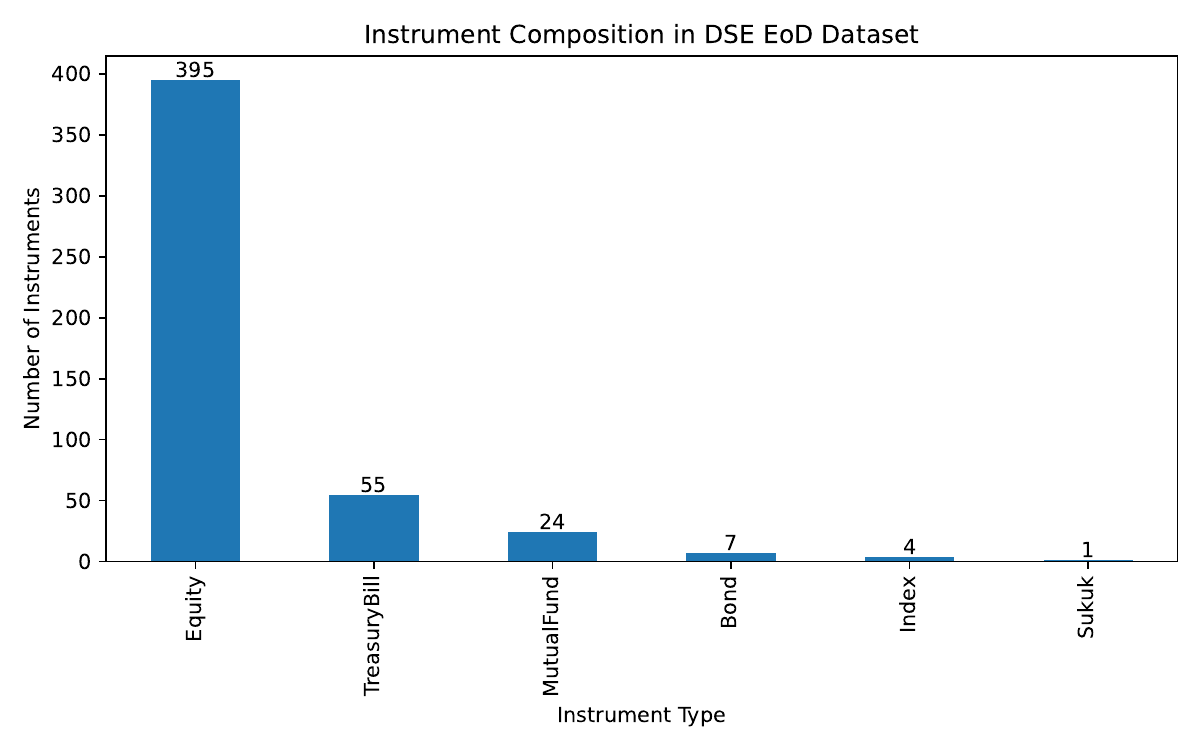}
  \caption{Instrument composition of the DSE end-of-day dataset by asset class. The dataset comprises 486 instruments across six categories, dominated by equities (395) and treasury bills (55). The multi-asset composition contributes to the heterogeneity of listing intervals observed in the panel.}
  \label{fig:D1}
\end{figure}

Figure~\ref{fig:D1} breaks down the dataset by instrument type. Equities account for the large majority (395 instruments), followed by treasury bills (55), mutual funds (24), bonds (7), indices (4), and a single sukuk. The multi-asset composition is relevant here: listing interval heterogeneity arises not just from staggered equity listings but also from the fact that entire instrument categories were introduced at different points in the exchange's history.

\subsection{Heterogeneous Listing Windows}

No two instruments in the dataset necessarily share the same observation window. Securities are listed at different times, meaning the point at which valid price data begins varies substantially across instruments. Each instrument's valid window is bounded by its first and last recorded trading date.

This creates a structural problem for panel construction. A calendar-aligned panel assigns every instrument a row for every date in the panel's range. If the alignment is done without tracking listing dates, instruments are implicitly assumed to have data for the full range --- and when they do not, preprocessing routines typically fill the gap. The result is artificial price observations before the instrument's actual listing date, which translate into near-zero returns that contaminate the return series.

\begin{figure}[htbp]
  \centering
  \includegraphics[width=0.75\linewidth]{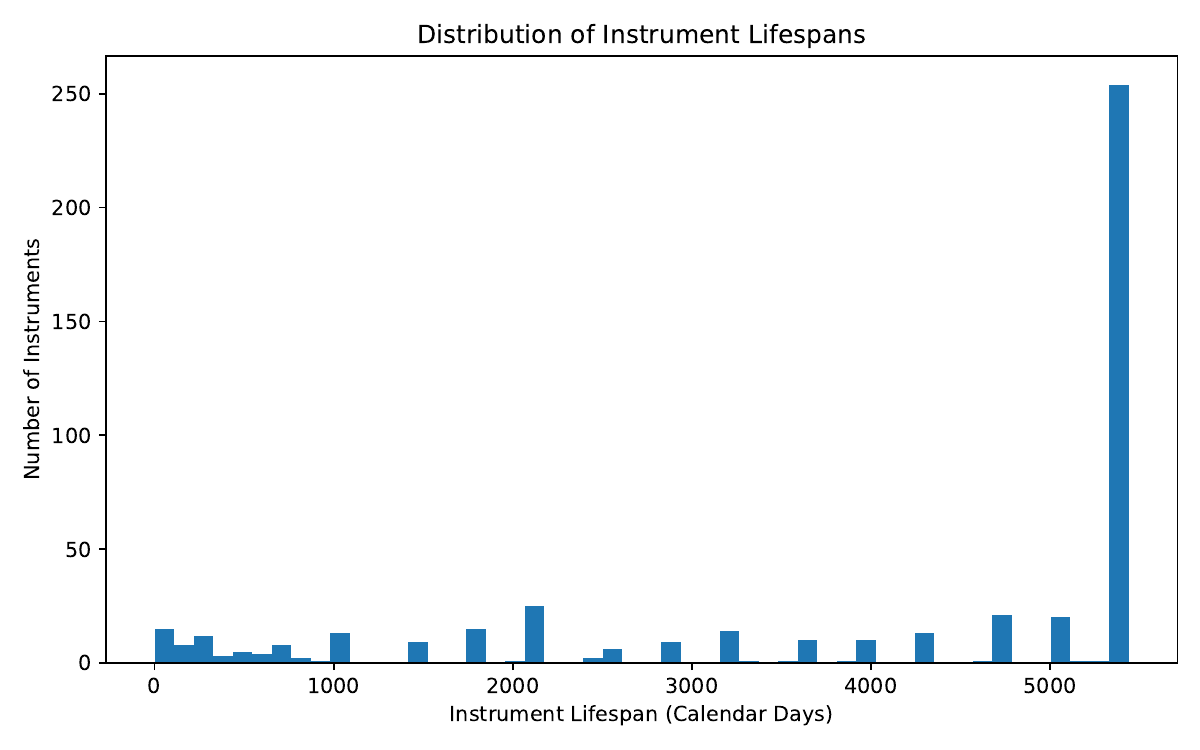}
  \caption{Distribution of instrument lifespans in calendar days. The spike at approximately 5,400 days corresponds to instruments listed since the start of the dataset observation window. The substantial proportion of instruments with lifespans below 2,000 days highlights the degree of listing heterogeneity and the extent of artificial backward padding that naive calendar alignment would introduce.}
  \label{fig:D2}
\end{figure}

Figure~\ref{fig:D2} shows the distribution of observed instrument lifespans in the dataset. The distribution is right-skewed, with a sharp spike at around 5,400 days marking instruments present since the start of the observation window. A large share of instruments falls below 2,000 days, meaning they were listed well after October 2012. These are the instruments most vulnerable to coverage distortion: under naive calendar alignment, their histories would be extended backward by several years, creating the longest artificial padding intervals.

\subsection{Coverage Encoding and Availability Matrix}

To handle this problem explicitly, the study introduces an availability matrix that records, for every instrument--date pair, whether valid observations exist. Let $S_i$ and $E_i$ denote the first and last observed trading dates for instrument $i$. The valid window is the interval $[S_i, E_i]$. The availability matrix $A_{i,t}$ is defined as

\begin{equation}
  A_{i,t} =
  \begin{cases}
    0 & \text{no valid observation available,} \\
    1 & \text{observation available in the adjusted dataset only,} \\
    2 & \text{observation available in the unadjusted dataset only,} \\
    3 & \text{observation available in both adjusted and unadjusted
          datasets.}
  \end{cases}
  \label{eq:availability}
\end{equation}

The four-value encoding captures not just presence or absence of data but also which dataset version carries valid observations on a given date. Unlike a balanced panel, which implicitly sets $A_{i,t} = 3$ everywhere, this representation preserves the actual trading participation pattern across instruments. Crucially, the availability matrix makes coverage a queryable property of the dataset itself. Any analysis that ingests this dataset - whether a volatility model, a clustering algorithm, a portfolio optimizer, or a machine learning pipeline - can condition its computation on $A_{i,t} > 0$ and thereby avoid the fabricated observations that naive alignment introduces. The correction requires no modification to the downstream model; it only requires that the data construction step is coverage-aware.

\begin{figure}[htbp]
  \centering
  \includegraphics[width=0.75\linewidth]{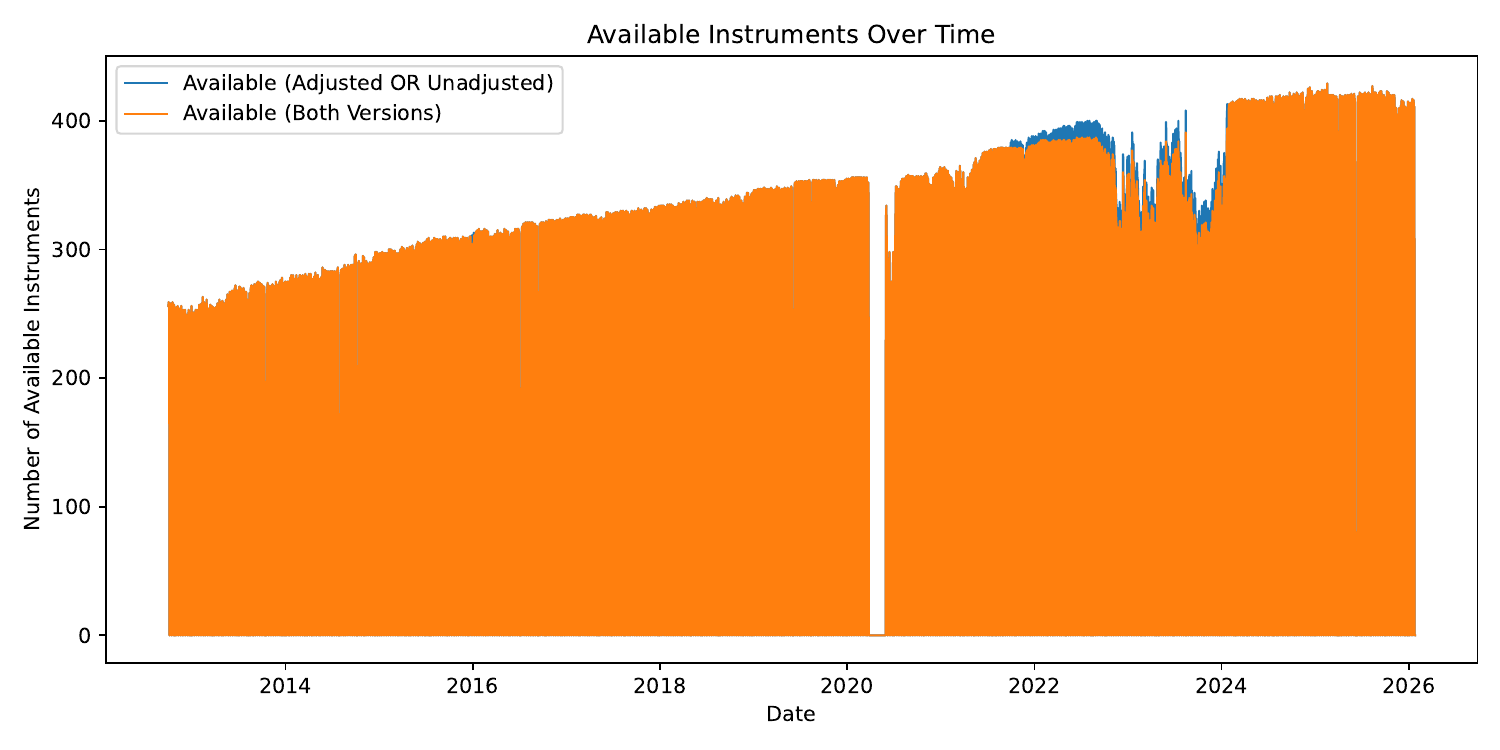}
  \caption{Number of available instruments per trading day across the dataset observation period (October 2012 -- January 2026). Orange indicates instruments available in both adjusted and unadjusted versions; blue indicates availability in at least one version. Instrument count grows from approximately 260 to 415, with panel coverage rising from 53\% to 85\% of the 486 total instruments. The drop in mid-2020 corresponds to the DSE market suspension during the COVID-19 pandemic. The concave dip around mid-2024 reflects market disruption during the July 2024 uprising in Bangladesh. Both are genuine market events preserved in the dataset without imputation.}
  \label{fig:C1}
\end{figure}

Figure~\ref{fig:C1} shows how the number of instruments with valid observations evolved over the dataset's timeline. The orange area counts instruments available in both adjusted and unadjusted versions; the blue trace counts those available in at least one version. The panel was never fully balanced: coverage started at around 260 instruments and grew to over 415 by January 2026, reflecting ongoing new listings throughout the period. Panel coverage over the full 486-instrument universe rose from roughly 53\% to 85\%. Two interruptions are visible in the series. The drop in mid-2020 corresponds to the DSE market suspension during the COVID-19 pandemic, when trading was halted for an extended period and valid observations fell sharply across all instruments simultaneously. The concave dip visible around mid-2024 reflects the political and market disruption surrounding the July 2024 uprising in Bangladesh, during which trading activity was severely disrupted. Both episodes are genuine market events rather than data collection failures, and they are preserved as-is in the dataset rather than imputed. A dataset constructed without the availability matrix would have no way to distinguish these legitimate gaps from synthetic pre-listing absences.

\subsection{Coverage-Aware Dataset Representation}

The complete dataset thus has three components:

\begin{enumerate}[label = \roman*)]
  \item \textbf{Instrument-level price series}, containing raw
    end-of-day trading observations.
  \item \textbf{Instrument metadata}, describing instrument
    characteristics and listing intervals.
  \item \textbf{Availability matrix}, encoding the temporal coverage
    of each instrument across the calendar.
\end{enumerate}

Together, these allow any downstream analysis to distinguish between dates where an instrument genuinely had no data and dates where valid data exist. The availability matrix is designed to be analysis-agnostic: the same dataset can support volatility estimation, pairwise distance computation, factor model construction, or machine learning fine-tuning, with each application querying $A_{i,t}$ to determine which observations are valid for its specific computation. Comparing results from a coverage-aware construction --- one that respects $[S_i, E_i]$ --- against naive constructions that extend or pad each series is the empirical strategy used throughout Section~\ref{sec:5}.

\section{Empirical Methodology}
\label{sec:4}

\subsection{Coverage-Aware and Naive Dataset Construction}

Two naive constructions of increasing severity are compared against the
coverage-aware baseline throughout the analysis. In the
\emph{coverage-aware} construction, statistical analysis for instrument
$i$ is confined to the interval $[S_i, E_i]$, using only dates on
which the instrument had valid observations.

The first naive construction, referred to as \emph{naive
forward-filled}, reindexes each instrument's price series to a
continuous calendar-day grid spanning from its first to its last
observed trading date, with forward-filling used to propagate the most
recent closing price across non-trading days. This mirrors the most
common practice in automated panel-building routines --- including
standard library functions such as \texttt{pandas.DataFrame.reindex}
with forward-fill, Bloomberg terminal panel exports, and many
commercial data vendor delivery formats --- where price series are
aligned to a continuous date range without restricting to confirmed
trading days. The result is a series that includes zero-return
observations on weekends, public holidays, and unscheduled market
closures --- dates on which no trading occurred and no valid price
was recorded.

The second naive construction, referred to as \emph{naive
backward-filled}, additionally extends each instrument's series
backward to the global panel start date by filling all pre-listing
dates with the first observed closing price. This represents the more
severe case where researchers apply backward imputation, use data
sources that silently pad pre-listing dates, or fill NaN values before
valid listing dates with the first available observation. The resulting
series contains a long segment of constant pre-listing prices that
translate directly into zero returns, more severely diluting the return
distribution than calendar-day alignment alone.

The comparison between estimates from these constructions and the
coverage-aware baseline quantifies coverage distortion at two levels of
severity.

\subsection{Return Computation}

Log returns are computed from closing prices throughout. For instrument
$i$ on date $t$:

\begin{equation}
  r_{i,t} = \ln(P_{i,t}) - \ln(P_{i,t-1})
  \label{eq:logreturns}
\end{equation}

where $P_{i,t}$ is the closing price. In the coverage-aware case,
returns are computed only within $[S_i, E_i]$, restricted to confirmed
trading days. In both naive cases, the padded price series introduces
zero-return observations --- on non-trading days in the forward-filled
construction, and additionally on all pre-listing dates in the
backward-filled construction --- which is where the distortion enters.

\subsection{Volatility Estimation Models}

Two estimation frameworks are used with distinct roles. ARIMA serves as
a mechanism illustration: it makes visible how zero-return padding
corrupts return dynamics and distorts model parameters. GARCH(1,1) is
the primary volatility analysis tool: its unconditional long-run
variance provides the main cross-instrument distortion measure, given
its sensitivity to distributional properties of the input series.

\subsubsection{ARIMA Modeling}

ARIMA models \cite{box2015time, siami2019comparative} are estimated on
the log return series for each instrument to illustrate how dataset
constructions affect model behavior. A fixed order of ARIMA(1,0,1) is
applied uniformly across all instruments in the cross-instrument
analysis for comparability; AIC and BIC are used for order selection
only in the single-instrument illustrative analysis of
Section~\ref{sec:5}. The ARIMA stage serves primarily as a mechanism
demonstration: zero-return observations introduced by either form of
naive alignment inflate the series length while suppressing measured
variability, which visibly distorts fitted parameters and forecast
accuracy metrics.

\subsubsection{GARCH Modeling}

The main volatility analysis uses GARCH(1,1)
\cite{bollerslev1986generalized}:

\begin{equation}
  \sigma_t^2 = \omega + \alpha r_{t-1}^2 + \beta \sigma_{t-1}^2
  \label{eq:garch}
\end{equation}

where $\omega > 0$, $\alpha, \beta \geq 0$, and $\sigma_t^2$ is the
conditional variance at time $t$. For cross-instrument comparison, the
unconditional long-run variance derived from the fitted GARCH(1,1)
parameters,

\begin{equation}
  \hat{\sigma}^2_\infty = \frac{\hat{\omega}}{1 - \hat{\alpha} -
  \hat{\beta}},
  \label{eq:unconditional_var}
\end{equation}

is used as the summary volatility measure, as it provides a single
comparable scalar per instrument. GARCH estimation is sensitive to the
distributional properties of the input series, making it a demanding
test for the effects of zero-return padding.

\subsection{Volatility Distortion Metrics}

Let $\sigma_{\text{aware}}$ and $\sigma_{\text{naive}}$ denote the
volatility estimates (return standard deviation or GARCH unconditional
long-run variance, respectively) from the coverage-aware and naive
constructions. The proportional distortion is:

\begin{equation}
  \Delta\sigma = \frac{\sigma_{\text{aware}} - \sigma_{\text{naive}}}
                      {\sigma_{\text{aware}}}
  \label{eq:distortion}
\end{equation}

A positive $\Delta\sigma$ means the naive construction produced a lower
volatility estimate than the coverage-aware one --- that is, the padding
suppressed the measured volatility. This is the expected direction of
the effect, since zero-return observations introduced by either form of
naive alignment pull the unconditional variance downward.

\subsection{Cross-Instrument Robustness Analysis}

The distortion metric is computed for each instrument in the sample and
the results are aggregated to assess consistency. If temporal coverage
bias is a real and systematic problem, $\Delta\sigma$ should be
positive and of meaningful magnitude across most instruments, not just
a handful of outliers.

\section{Empirical Results}
\label{sec:5}

\subsection{Illustrative Single-Instrument Example}

\begin{figure}[htbp]
  \centering
  \includegraphics[width=0.80\linewidth]{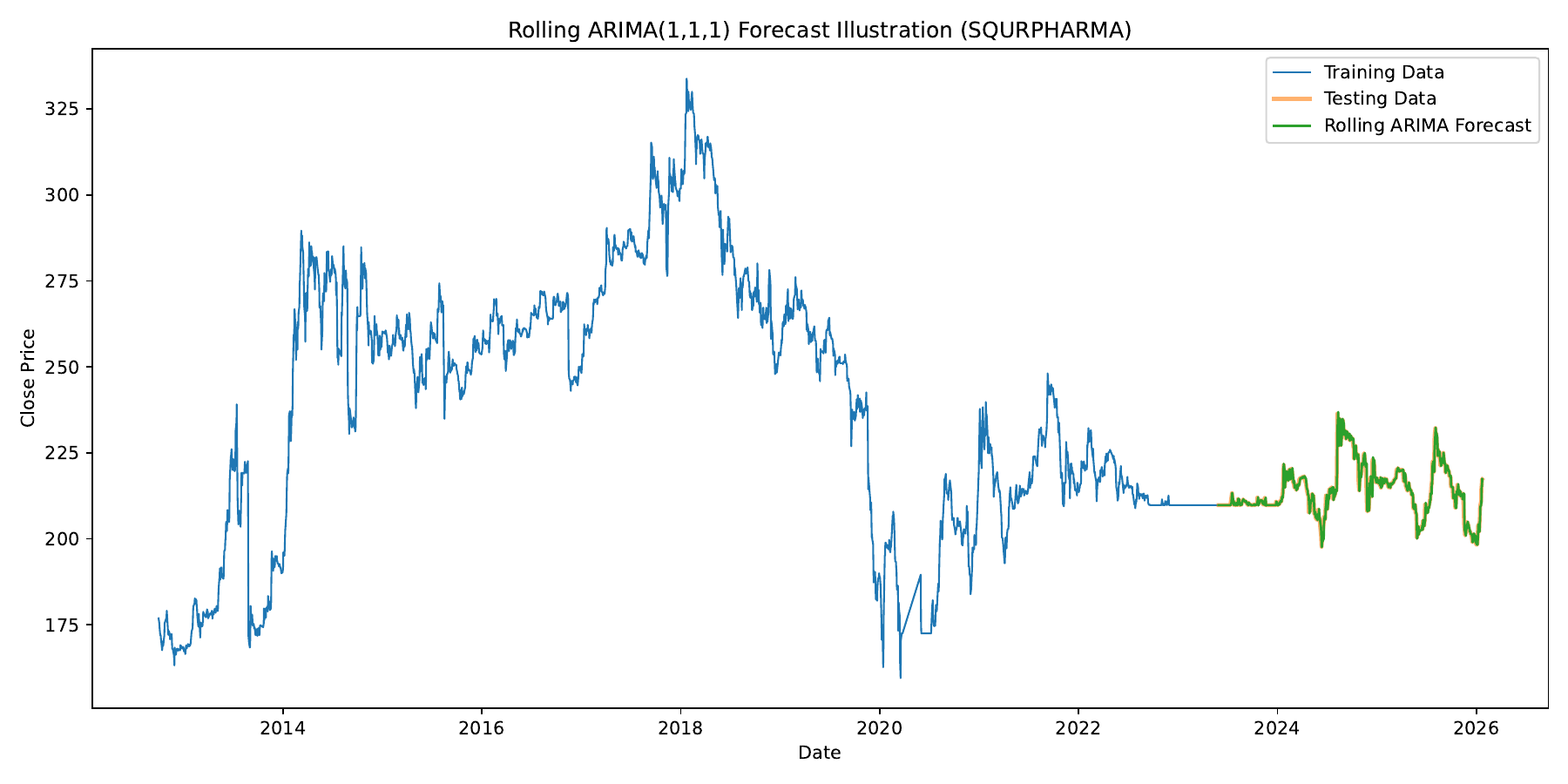}
  \caption{Rolling ARIMA(1,1,1) forecast illustration for SQURPHARMA
    under the coverage-aware construction. Blue: training data over
    the valid listing interval; orange: test data (overlaps closely
    with training data at this scale); green: rolling one-step-ahead
    forecast. SQURPHARMA predates the panel start so coverage bias
    does not affect this specific instrument; the figure illustrates
    the forecast methodology under a clean coverage-aware
    construction. For instruments listed after 2016, naive calendar
    alignment would pad the training window with zero-return
    observations on non-trading days and, under backward extension,
    on pre-listing dates as well, suppressing return variance and
    distorting model estimates.}
  \label{fig:A1}
\end{figure}

The analysis begins with a concrete single-instrument illustration
before moving to the cross-instrument results. Figure~\ref{fig:A1}
uses SQURPHARMA to illustrate the forecasting methodology under the
coverage-aware construction. SQURPHARMA has been listed on the DSE
since October 2012, giving it a valid observation window that spans
nearly the full dataset period; under either naive construction,
nothing changes for this particular instrument since its listing
predates the panel start. The figure is therefore not a distortion
demonstration for SQURPHARMA itself, but a reference illustration
of what a clean coverage-aware forecast looks like. For instruments
listed after 2016, the same forecast setup would operate on a return
series contaminated by zero-return observations --- on non-trading days
through forward-filling, and on pre-listing dates through backward
extension --- compressing measured variance and biasing model
parameters before a single genuine return is observed.

Table~\ref{tab:coverage_vs_naive_comparison} presents a direct
comparison using AAMRANET, an instrument listed substantially after
the panel start and therefore directly affected by coverage bias. The
coverage-aware construction yields 1,954 observations confined to
confirmed trading days. The naive forward-filled construction expands
this to 3,034 observations by including weekends, public holidays, and
unscheduled market closures within the instrument's observed history
--- a difference of 1,080 calendar-day observations carrying zero
returns. Return standard deviation falls from 0.02619 to 0.02102, a
suppression of 19.8\%. The naive backward-filled construction extends
the series further to 4,861 observations by additionally prepending
all pre-listing dates back to the global panel start with the first
observed closing price, introducing a further 1,827 pre-listing
observations with constant prices and zero returns. Return standard
deviation falls further to 0.01658, a suppression of 36.6\%. In both
naive cases, RMSE and MAE decline alongside standard deviation, but
this reflects dilution of genuine return variability rather than
improved forecasting accuracy. Applying the backward-filled
construction across the full 53-instrument sample caused GARCH
non-convergence in 22 instruments ($\hat{\alpha} + \hat{\beta} \geq
0.999$), indicating that pre-listing backward extension corrupts the
return distribution severely enough to break conditional variance
estimation entirely --- a finding that itself underscores the gravity
of the bias. The cross-instrument robustness analysis in
Section~\ref{sec:5.2} therefore uses the forward-filled construction,
which represents the more common preprocessing pattern and admits
stable estimation across all 53 instruments. The 26\% GARCH distortion
reported there should consequently be understood as a lower bound on
the full severity of the effect: the backward-filled construction ---
which broke GARCH entirely in 41\% of instruments --- represents a
more extreme failure mode that the forward-filled analysis is
deliberately designed to understate.

\begin{table}[htbp]
  \caption{ARIMA(1,0,1) model statistics for AAMRANET under
    coverage-aware and two naive dataset constructions. The naive
    forward-filled construction introduces 1,080 additional
    calendar-day observations through alignment without trading-day
    restriction, suppressing return standard deviation by 19.8\%.
    The naive backward-filled construction additionally extends the
    series to the global panel start date, introducing a further
    1,827 pre-listing observations and suppressing return standard
    deviation by 36.6\%. Both naive constructions reduce apparent
    forecast error metrics through dilution of genuine return
    variability rather than improved model fit.}
  \label{tab:coverage_vs_naive_comparison}
  \centering
  \begin{tabular}{lcccccc}
    \toprule
    \textbf{Model} & \textbf{Obs.} & \textbf{Return STD}
    & \textbf{AIC} & \textbf{BIC} & \textbf{RMSE} & \textbf{MAE} \\
    \midrule
    Coverage-Aware        & 1954 & 0.02619 & $-$6839.88  & $-$6818.47
    & 0.02221 & 0.01615 \\
    Naive Forward-Filled  & 3034 & 0.02102 & $-$11692.46 & $-$11669.28
    & 0.01788 & 0.01067 \\
    Naive Backward-Filled & 4861 & 0.01658 & $-$21022.86 & $-$20997.79
    & 0.01801 & 0.01080 \\
    \bottomrule
  \end{tabular}
\end{table}

\subsection{Cross-Instrument Volatility Distortion}
\label{sec:5.2}

To assess how general this distortion is, the methodology from
Section~\ref{sec:4} is applied across a set of instruments with
heterogeneous listing histories. Instrument selection applied two
criteria:

\begin{enumerate}[label = \roman*)]
    \item First observed on the DSE after 2016
    \item At least 400 trading days of observations
\end{enumerate}

Instruments satisfying both criteria were therefore listed more than
four years after the start of the observation window, ensuring a
meaningful minimum padding interval under naive calendar alignment
while retaining sufficient trading history for stable model
estimation. The minimum of 400 trading days was imposed to ensure
sufficient history for stable ARIMA and GARCH parameter estimation
rather than as a data quality filter; this criterion necessarily
excludes the shortest-history instruments --- those most severely
affected by coverage bias --- and therefore produces a conservative
lower-bound estimate of the true distortion magnitude. This yields a
sample of 53 instruments across different listing years and
observation windows.

For each instrument, return volatility is estimated under the
coverage-aware and naive forward-filled constructions, and the
distortion metric from Equation~\eqref{eq:distortion} is computed. The
cross-instrument analysis uses forward-filled only, since the
backward-filled construction caused GARCH estimation failure in 22 of
53 instruments; the reported 26\% GARCH distortion therefore
understates the full severity of the bias and should be interpreted
accordingly. The results are consistent. Naive temporal alignment
reduces estimated return volatility in every instrument in the sample,
with a mean distortion of 20.0\% and a median of 19.7\%. Every
instrument exceeds 10\% suppression, and more than a third exceed 20\%.
The picture is starker for GARCH: mean unconditional variance
distortion exceeds 26\%, and the effect is present in over 90\% of
the instruments. A one-sample sign test on the direction of return STD
distortion confirms the result is not due to chance: all 53 instruments
show positive distortion ($p < 0.0001$). A one-sample $t$-test on the
mean distortion rejects zero mean ($t = 26.6$, $p < 0.0001$),
confirming the suppression effect is statistically significant across
the sample.

\subsection{Distribution of Volatility Distortion}

\begin{figure}[htbp]
  \centering
  \includegraphics[width=0.65\linewidth]{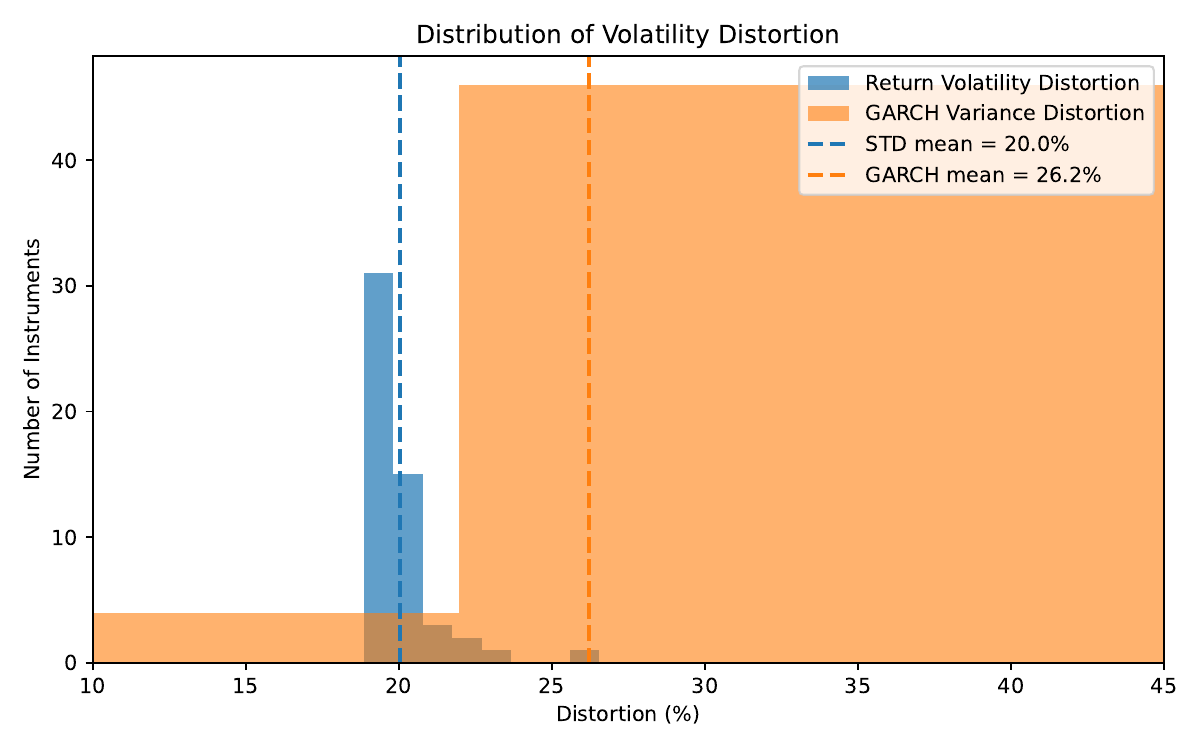}
  \caption{Empirical distribution of volatility distortion across 53
    instruments under the naive forward-filled construction. Return
    volatility distortion values (blue) are concentrated near 20\%,
    while GARCH unconditional variance distortions (orange) exhibit a
    wider distribution with a higher mean of 26.2\%, confirming that
    naive temporal alignment systematically biases volatility
    estimation downward. Both distributions reflect forward-filling
    only; the backward-filled construction produced GARCH
    non-convergence in 41\% of instruments, representing a more
    severe failure mode.}
  \label{fig:V1}
\end{figure}

Figure~\ref{fig:V1} shows the distribution of distortion values across
the 53 instruments for both return volatility and GARCH unconditional
variance. The return STD distribution (blue) is tightly concentrated
around 20\%, with little spread. The GARCH distribution (orange) is
centered higher, around 26\%, but is considerably wider. In both cases,
the mass of the distribution is clearly positive, confirming that the
naive construction suppresses volatility estimates rather than inflating
them.

\begin{figure}[htbp]
  \centering
  \includegraphics[width=0.65\linewidth]{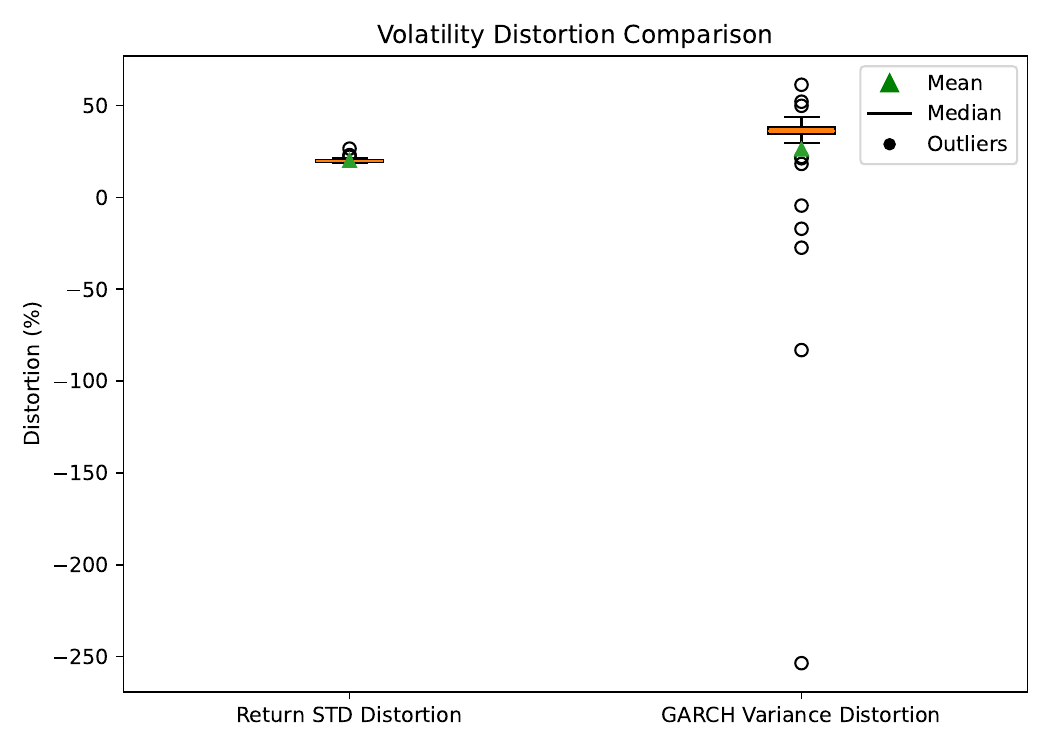}
  \caption{Boxplot comparison of volatility distortion distributions
    across return standard deviation and GARCH unconditional variance
    estimates under the naive forward-filled construction. The extreme
    negative outliers in GARCH distortion correspond to instruments
    where the non-trading-day padding interval is disproportionately
    long relative to the valid observation window, causing nonlinear
    divergence in unconditional variance estimates. These instruments
    tend to be the most recently listed with the highest
    padding-to-valid-observation ratios; under the backward-filled
    construction, analogous cases caused outright GARCH
    non-convergence in 22 of 53 instruments.}
  \label{fig:V2}
\end{figure}

Figure~\ref{fig:V2} shows the same data as boxplots. The return STD
column is compact, with a narrow interquartile range and no major
outliers. The GARCH column is wider, and a cluster of extreme negative
outliers is visible below zero. These correspond to instruments where
the artificial padding interval is long enough relative to the valid
observation window that the GARCH variance process converges to a
qualitatively different solution under the naive construction --- the
distortion flips direction rather than simply amplifying. That these
cases exist reinforces rather than undermines the main finding: GARCH
estimation is sensitive to series composition, and synthetic
non-trading-day observations can push it in unexpected directions.

\subsection{Mechanism Analysis}

\begin{figure}[htbp]
  \centering
  \includegraphics[width=0.65\linewidth]{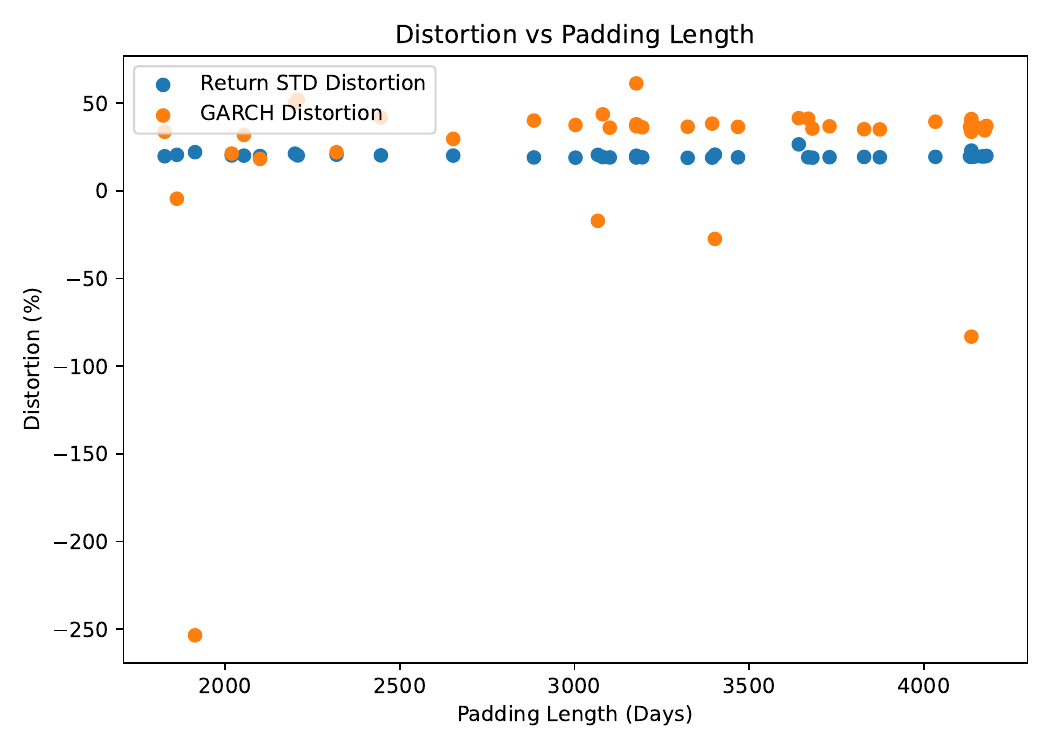}
  \caption{Volatility distortion as a function of padding length (days)
    for return standard deviation (blue) and GARCH unconditional
    variance (orange) under the naive forward-filled construction.
    Return STD distortion is tightly clustered near 20\% across all
    listing vintages, consistent with a roughly constant ratio of
    non-trading to total calendar days across instruments. GARCH
    unconditional variance distortion shows wider dispersion and a
    broadly positive association with longer padding length. Outliers
    with large negative GARCH distortion correspond to instruments
    where the naive construction causes the GARCH variance process to
    converge to a qualitatively different solution; these are
    instruments with the most extreme padding-to-valid-observation
    ratios.}
  \label{fig:V4}
\end{figure}

The central question in this section is whether padding length --- the
number of non-trading-day observations introduced by naive calendar
alignment --- is what drives the distortion. Figure~\ref{fig:V4} plots
$\Delta\sigma$ against padding length for all 53 instruments.

For return STD distortion (blue), the answer is clear: the points
cluster tightly near 20\% regardless of listing vintage. This is
consistent with the mechanics of calendar-day alignment: the ratio of
non-trading days to total calendar days is approximately constant
across all instruments (roughly one in three calendar days is a
non-trading day), so the dilution of unconditional variance is
similarly constant regardless of when the instrument was listed.

GARCH distortion (orange) tells a more complicated story. The bulk of
the points sit above 20\% and show a weak positive association with
padding length, but the spread is wide and the negative outliers are
scattered rather than concentrated at long padding lengths. This
reflects the nonlinear dynamics of GARCH estimation: conditional
variance models can converge to different local solutions depending on
the composition of the input series, and that sensitivity is not simply
a function of padding length alone. The main conclusion stands ---
padding drives the distortion --- but GARCH introduces
instrument-level heterogeneity that return STD does not.

Taken together, the results in this section confirm that temporal
coverage assumptions are not a minor technicality. When instrument
listing windows are ignored, the resulting distortions are large,
consistent, and in a predictable direction --- systematically downward
for return STD, and predominantly downward with occasional nonlinear
exceptions for GARCH.

\section{Conclusion}
\label{sec:6}

This paper has examined what happens to volatility estimates when
financial panels are constructed without accounting for heterogeneous
instrument listing intervals. The problem is not hypothetical: across
53 DSE instruments selected for having meaningful padding intervals
under naive alignment, return volatility was suppressed by an average
of 20\% and GARCH unconditional variance by over 26\% under
calendar-day forward-filling alone --- the milder of the two naive
constructions studied, and therefore a lower bound on the full
severity. The single-instrument illustration demonstrates that the
severity escalates with backward extension: applying backward-fill to
pre-listing dates suppresses return volatility by 36.6\% for the same
instrument, and when applied across all 53 instruments causes GARCH
estimation to break down entirely in 22 cases (41\%). The direction is
consistent across both naive constructions --- the naive alignment
always produces lower volatility --- and the mechanism is
straightforward: zero-return observations introduced by padding,
whether on non-trading days or pre-listing dates, dilute unconditional
variance and bias conditional variance models that process the full
series.

The coverage-aware framework introduced here addresses this by making
listing intervals a formal part of the dataset structure. An
availability matrix records which instruments have valid observations
on which dates, and statistical analysis is restricted accordingly.
This is not a complex fix, but it requires treating coverage as an
explicit modeling input rather than an implicit preprocessing detail.

The practical stakes are real. A practitioner who builds a panel
without this correction will systematically underestimate the
volatility of recently listed instruments --- producing understated
Value at Risk (VaR) figures, mispriced options, and portfolio
allocations that underweight risk for newer securities. For
researchers, the implication is that empirical results in volatility
modeling and asset pricing may be sensitive to how their underlying
datasets were constructed, particularly if those datasets include
instruments with heterogeneous listing histories.

The DSE setting used here was chosen for its rich diversity of listing
vintages and instrument types, but the problem is not specific to any
single market or analytical framework. The following analytical
contexts are all susceptible to temporal coverage bias whenever a
calendar-aligned panel is constructed without explicit coverage
encoding:

\begin{itemize}
  \item \textbf{Volatility and risk estimation}: as demonstrated in
    this paper, GARCH-based conditional variance and unconditional
    return variance are both suppressed, producing understated VaR,
    mispriced derivatives, and biased covariance matrices for
    portfolio construction.

  \item \textbf{Pairwise behavioral similarity analysis}: DTW
    distance matrices, Pearson correlation matrices, and any other
    pairwise measure computed over a calendar-aligned panel will
    encode phantom similarity among recently-listed instruments and
    phantom dissimilarity between instruments with different listing
    ages.

  \item \textbf{Factor models and cross-sectional asset pricing}:
    Fama--French style regressions and factor loading estimates
    operate on the full panel return matrix; fabricated pre-listing
    observations bias both factor exposures and residual variance
    estimates.

  \item \textbf{Cointegration and long-run relationship testing}:
    spurious stationarity in the padding segment can mask or
    manufacture cointegrating relationships between instruments,
    leading to incorrect inference about long-run dynamics.

  \item \textbf{Backtesting and event study methodology}: any
    performance backtest or event study window that draws from a
    calendar-aligned panel without coverage encoding silently
    includes pre-listing periods as valid observations, inflating
    apparent sample sizes and distorting pre-event baselines.

  \item \textbf{Machine learning and temporal foundation model
    fine-tuning}: models trained or fine-tuned on panel-formatted
    data will learn from fabricated observations as genuine signal;
    temporal foundation models that process multiple instruments
    on a common time axis are particularly susceptible, as
    coverage-biased panels are expected to degrade embedding
    quality in proportion to the fraction of recently-listed
    instruments.
\end{itemize}

The correction is identical across all of these settings: restrict
each instrument's analytical window to $[S_i, E_i]$ as encoded in
the availability matrix. The framework and dataset introduced here
make this restriction a default property of the data rather than a
case-by-case modelling decision, enabling any downstream analysis
to adopt coverage-aware computation without changes to its own
methodology.

Beyond financial markets, the coverage-aware framework has direct
applicability in any domain where a panel combines entities with
heterogeneous observation start dates. The structural problem is
identical: entities are aligned to a common calendar without
tracking when each entity's valid observations began, and any
analysis that treats the resulting panel as ground truth inherits
the distortion. Concrete non-financial settings where this arises
include: IoT and sensor networks, where nodes or stations are
commissioned incrementally and a network-level panel will contain
different proportions of valid versus padded observations for each
node depending on its deployment date; longitudinal clinical and
epidemiological datasets, where patients or subjects enroll at
different times and pre-enrollment imputation distorts baseline
measurements and treatment effect estimates; country-level
macroeconomic panels (World Bank, IMF), where newly tracked
indicators or newly included countries have shorter histories than
long-standing ones; environmental and climate monitoring arrays,
where stations are installed over time and satellite or sensor
coverage grows progressively; and social media and platform
analytics panels, where accounts, products, or content entities
have inception dates that vary widely within the observation window.
In every case, the availability matrix formulation from
Equation~\eqref{eq:availability} generalises directly: replace
``instrument'' with ``entity,'' replace ``listing date'' with
``entry date'' or ``inception date,'' and the coverage-aware
restriction $[S_i, E_i]$ applies without modification. Extending
empirical validation of the coverage-aware framework to these
domains is a natural and consequential direction for future work.


\section*{Declarations}

\paragraph{Availability of data and materials.}
The dataset supporting the conclusions of this article is publicly
available on Mendeley Data at
\url{https://doi.org/10.17632/23553sm4tn.4}~\cite{muhammad2026dhaka}.
The analysis code is available on GitHub at
\url{https://github.com/TashreefMuhammad/Dhaka-Stock-Exchange-EoD-Dataset-Metadata}.

\paragraph{Competing interests.}
The author declares no competing interests.

\paragraph{Funding.}
This research received no specific grant from any funding agency in
the public, commercial, or not-for-profit sectors.

\paragraph{Author's contributions.}
T.M. conceived the study, curated the dataset, developed the
coverage-aware structuring framework, conducted all empirical analyses,
and wrote the manuscript.

\paragraph{Acknowledgements.}
Earlier versions of the dataset were developed in collaboration with
Rakibul Islam and Mohammad Shafiul Alam, whose prior contributions are
gratefully acknowledged.


\bibliographystyle{unsrt}
\bibliography{references}

\end{document}